\documentclass[aps,showpacs]{revtex4}
\usepackage{amsbsy,latexsym}
\usepackage{amsmath}
\usepackage{amssymb, mathrsfs}
\usepackage[mathscr]{eucal}

\def\d{\operatorname{d}}\def\<{\langle}\def\>{\rangle}
\def\Tr{\operatorname{Tr}}\def\:{\hbox{\bf :}}
\def\spc#1{\mathcal{#1}}
\def\openone{1\!\!1}
\def\Reals{\mathbb R}\def\Cmplx{\mathbb C}
\def\Bndd#1{\alg{B}(#1)}

\def\alg#1{{\mathcal #1}}

\def\qed{$\,\blacksquare$\par}

\def\geq{\geqslant}
 
\newtheorem{Lemma}{Lemma}

\newtheorem{Theo}{Theorem}
\def\Proof{\medskip\par\noindent{\bf Proof. }}

\def\>{\rangle}
\def\<{\langle}

\begin{document}

\title{Realization of continuous-outcome measurements on finite dimensional quantum systems}

\author{G.~Chiribella}
\author{G.~M.~D'Ariano}
\author{D.-M.~Schlingemann}

\affiliation{Dipartimento di Fisica A. Volta, Universita di Pavia}
\date{\today}

\begin{abstract}
This note contains the complete mathematical proof of the main Theorem of the paper  ``How continuous measurements in finite dimension are actually discrete" \cite{PRL}, thus showing that in finite dimension any measurement with continuous set of outcomes can be simply realized by randomizing some classical parameter and conditionally performing a measurement with  finite outcomes.   
\end{abstract}

\maketitle

In the following we prove that any quantum measurement performed on a finite dimensional system can be realized by randomly drawing the value of a suitable classical parameter, and conditionally performing a quantum measurement with finite outcomes and a suitable post-processing operation.

\section{The convex set of quantum measurements with given outcome space}

\subsection{POVMs in finite dimensional systems}

Consider an experiment where a quantum system is measured, producing
an outcome $\omega$ in the outcome space $\Omega$. The possible events
in the experiment can be identified with subsets of $\Omega$, the
subset $B \subseteq \Omega$ corresponding to the event ``the
measurement outcome $\omega$ is in $B$''.  More precisely, the
probability distribution of the possible outcomes is specified by by
the probabilities $p(B)$ of events in a sigma-algebra $\sigma
(\Omega)$ of measurable subsets of $\Omega$. As it happens in all physical
situations, we will assume here the outcome space $\Omega$ to be a
Hausdorff space, namely a topological space where for any couple of
points $\omega_1, \omega_2 \in \Omega$ there are two disjoint open
sets $U_1, U_2 \subset \Omega, \quad U_1 \cap U_2 = \emptyset$ such
that $\omega_1 \in U_1$ and $\omega_2 \in U_2$.  Accordingly, the
sigma-algebra $\sigma (\Omega)$ will be the Borel sigma-algebra,
generated by intersections and countable unions of open sets.

Let $\spc H$ be the Hilbert space of a $d$-level quantum system, and $\Bndd {\spc H}$  the set of linear operators on $\spc H$. Let  $P: \sigma (\Omega) \to \Bndd{\spc H}$  be a POVM with outcome space $\Omega$, $\sigma (\Omega)$ denoting the sigma-algebra of all measurable subsets in $\Omega$. By definition, the POVM $P$ associates a measurable subset $B \in \sigma (\Omega)$ with an operator $P(B) \in \Bndd{\spc H}$, satisfying the conditions 
\begin{enumerate}
\item  $P(B) \ge 0 \quad \forall  B \in \sigma (\Omega)$ (positivity)
\item  $P(\Omega) = \openone$ (normalization) 
\item  $P(\cup_k B_k) = \sum_k P(B_k)$ for mutually disjoint $\{B_k\}$ (countable additivity).
\end{enumerate}  
In the following we will fix both the finite-dimensional Hilbert space $\spc H$ and the outcome space $\Omega$. We will denote with $\mathscr{P}$ the convex set of
all POVMs for $\spc H$ with outcome space $\Omega$.  

The following Lemma, proved in Ref. \cite{PRL}, provides a convenient
way of representing POVMs:
\begin{Lemma}[Existence of a POVM density] \label{Lemma:Density}
  Every POVM $P\in\mathscr P$ admits a {\em density} with unit trace,
  namely for any POVM $P$ there exists a finite measure $\mu
  (\d\omega)$ over $\Omega$ such that $\mu(\Omega) =d$ and
\begin{equation}\label{POVMDensity}
P(B) = \int_{B} \mu (\d \omega)~ M(\omega)~,
\end{equation}
with $M(\omega)\geq 0$ and $\Tr[M(\omega)]=1$ $\mu$-almost everywhere.
\end{Lemma}

\subsection{Topological properties of the set of POVMs}
Given a quantum state $\rho$ on $\spc H$, the POVM density $M(\omega)$ provides the probability density of the measurement outcome $\omega \in \Omega$ via the Born rule $p(\omega) = \Tr[\rho M(\omega)]$. Once the probability distribution  has been specified one can take the expectation value of any continuous function $f: \Omega \to \Reals$, given by
\begin{equation}\label{Expectation}
\mathbb E_{\rho}(f) = \int_{\Omega} \mu (\d \omega)~f(\omega) \Tr[\rho M(\omega)]~.
\end{equation}
Note that the POVM $P$ is completely specified by the expectation
values that it provides for arbitrary functions $f(\omega)$ and
arbitrary states $\rho$.

Let $\mathcal V$ be the vector space of all operator-valued functions $A: \Omega \to
\Bndd{\spc H}$ that are continuous and bounded with respect to the
trace norm in $\Bndd{\spc H}$.  The space $\mathcal V$ is a Banach space equipped
with the norm 
\begin{equation}
|| A|| := d ~ \sup_{\omega \in \Omega} ||A(\omega)||_1~
\end{equation}
(here we are rescaling by $d$ the sup of the trace norm just for later convenience),
and its dual space $\mathcal V^*$ is the Banach space of all linear
functionals $\lambda: \mathcal V \to \Cmplx$ that are bounded with
respect to the dual norm $|| \lambda|| =: \sup_{A \in \mathcal V,
||A|| =1}~ | \lambda (A) |$.

Then we have the following 
\begin{Lemma}[Embedding of POVMs in a dual Banach space]\label{Lemma:BallSubset}
The set of POVMs $\mathscr P$ can be embedded in a convex subset $\mathscr C$ of the unit ball in the dual Banach space $\mathcal V^*$. Such a convex subset $\mathscr C \subseteq \mathcal V^*$ is defined as the set of all functionals $\lambda \in \mathcal V^*$ satisfying the two properties:
\begin{enumerate}
\item for any positive $A \in \mathcal V$ (i.e. $A(\omega) \ge 0 \quad \forall \omega \in \Omega$), $\lambda (A)  \ge 0$ (positivity)
\item for any constant $A \in \mathcal V$ (i.e. of the form $A (\omega) 
= A_0$ for some fixed operator $A_0$), $\lambda (A) = \Tr [A_0]$ (normalization)
\end{enumerate}
The correspondence $\mathscr P \to \mathscr C$ is injective. 
\end{Lemma}
\Proof First, we prove that the set $\mathscr C$ defined in the Lemma is actually a subset of the unit sphere in $\mathcal V^*$.   Indeed, for any $\lambda \in \mathscr C$ and for any $A \in \mathcal V$ one has
\begin{eqnarray}
|\lambda (A)| &=& |\lambda (A_+) - \lambda (A_-)|\\
&\le& |\lambda (A_+)| + |\lambda (A_-)|\\
&=& \lambda (|A|)\\
&\le& \lambda (\openone)~ \sup_{\omega \in \Omega} ||A(\omega)|| \\
&=& d ~\sup_{\omega \in \Omega} ||A(\omega)||\\
&=& ||A||~,
\end{eqnarray}
where we defined $A_+\ge 0$ ($A_-\ge 0$) and $|A| \ge 0$ as the positive (negative) part and the the modulus of $A$, respectively, i.e. $A(\omega) = A_+ (\omega) - A_-(\omega)$ and $|A (\omega)| = A_+ (\omega) + A_-(\omega)$. The second inequality comes from the positivity of $\lambda$ and from the inequality $|A (\omega)| \le \openone ~\sup_{\omega \in \Omega} ||A(\omega)||_1 $. The third equality comes from the normalization of $\lambda$ on the constant function $\openone$.  Since we have $|\lambda (A)| \le ||A ||$ for any $A \in \mathcal V$, necessarily $||\lambda|| \le 1$. 

Now we prove the embedding. For any POVM $P \in \mathscr P$ take the linear functional $\lambda_P$ defined by 
\begin{equation}
\lambda_P (A) = \int_{\Omega} \mu (\d \omega) ~ \Tr[A(\omega) M(\omega)]~,
\end{equation} 
where $\mu (\d \omega)$ and $M(\omega)$ are the scalar measure and the
density associated to $P$ as in Lemma \ref{Lemma:Density},
respectively. The functional $\lambda_P$ is in $\mathscr C$, since the
positivity and the normalization of $\lambda_P$ are a direct
consequence of the positivity and the normalization of the POVM $P$. 
 
Finally, the correspondence $P \to \lambda_P$ is injective, since the
functional $\lambda_P$ specifies the expectation value produced by the
POVM $P$ for any function $f(\omega)$ and for any quantum state
$\rho$, according to $\mathbb E_{\rho} (f) = \lambda_{P} (f
\rho)$. Hence, if $\lambda_{P_1} = \lambda_{P_2}$, then $P_1$ and
$P_2$ have the same expectation values, namely $P_1=P_2$. \qed

The previous Lemma associates any POVM $P$ with a functional
$\lambda_P \in \mathcal V^*$ in an invertible fashion.  
Moreover, in the case of compact outcome space $\Omega$ one can prove also the converse: 
\begin{Lemma}[One-to-one correspondence for compact outcome spaces]\label{Lemma:IdentificationCompact}
For compact $\Omega$, the set of all POVMs $\mathscr P$ is in one-to-one correspondence with the set $\mathscr C \subseteq \mathcal V^*$ defined in Lemma \ref{Lemma:BallSubset}.
\end{Lemma}

\Proof We only have the show that the correspondence $P \to \lambda_P$ is surjective. To this purpose, let $\lambda$ be a functional in $\mathscr C$. Define the POVM $P$ by the expectation values $\mathbb E_{\rho} (f): = \lambda (f \rho)$, for any function $f(\omega)$ and any quantum state $\rho$. The positivity and the normalization of the POVM $P$ follow directly from the positivity and the normalization of the functional $\lambda$. To prove sigma-additivity, fix a quantum state $\rho$, and consider the expectation $\mathbb  E_{\rho}$ as a functional on  $\mathcal C (\Omega)$.  Since for compact $\Omega$ the dual Banach space of  $\mathcal C(\Omega)$ is the space of probability measures over $\Omega$,  the functional $\mathbb E_{\rho}$ can be representated as the average of $f$ with respect to a probability measure $\mu_{\rho}$. The sigma-additivity of the POVM $P$ then follows from the sigma-additivity of $\mu_{\rho}$ for any $\rho$. \qed 
{\bf Remark (noncompact outcome spaces and sigma-additivity).}  If
$\Omega$ is noncompact, then the set $\mathscr C$ contains also
functionals that do not correspond to any POVM in $\mathscr
P$. Essentially, these functionals correspond to the possible ways to
take limits for $\omega$ going to infinity. More precisely, the action
of a functional of this kind on an element $A \in \mathcal V$ is given
by 
\begin{equation}
\lambda (A) =
\lim_{\gamma \in \Gamma} \Tr\left[A(\omega_{\gamma})\right]~,
\end{equation}
where $\Gamma$ is a directed set and $\{\omega_{\gamma}~|~ \gamma \in
\Gamma\}$ is a net of points in $\Omega$   with the property that for any compact set $B \subset \Omega$ there exists a $\gamma_0 \in \Gamma$ such that for any $\gamma \ge \gamma_0$ one has $\omega_{\gamma} \not \in B$.      It is easy to see that all these
functionals satisfy the positivity and the normalization,
i.e. $\lambda (A) \ge 0$ for positive $A$ and $\lambda (A_0) =
\Tr[A_0]$ for constant $A_0$. However, in these cases the POVM
$P$ defined by $\mathbb E_{\rho}(f) = \lambda (\rho f)$ would  not be
sigma-additive.  Indeed one would have $P(B)=0$ for any compact set $B$, while
$P(\Omega) = \openone$.  For a set of compact sets $\{B_k\}$ such that $\cup_k B_k = \Omega$ one would have $0 = \sum_k P(B_k) < \openone = P(\Omega)$, thus violating sigma-additivity.

Thanks to Lemma \ref{Lemma:BallSubset}, the set $\mathscr P$ of all
POVMs can be injectively embedded in a subset of the unit ball in the
dual space $\mathcal V^*$.  The following Lemma \ref{Lemma:Closed} and
Theorem \ref{Theo:Compact} present the topological properties of the
embedding set $\mathscr C$.
\begin{Lemma}[Closure of $\mathscr C$]\label{Lemma:Closed}
The embedding set $\mathscr C$  is closed in the weak-* topology.
\end{Lemma}   
\Proof  Let  $\Gamma$ be a directed set and $\{\lambda_{\gamma} \in \mathscr C~|~ \gamma \in \Gamma \}$ a net that converges to a linear functional $\lambda \in \mathcal V^*$ in the weak-* topology. By definition, this means that  $\lim_{\gamma}  \lambda_{\gamma}(A)=\lambda(A)$ for all $A \in \mathcal V$. In particular,  one has  \emph{i)} for any positive $A \in \mathcal V$, $\lambda(A) = \lim_{\gamma} \lambda_{\gamma}(A) \geq 0$ and \emph{ii)}  for any constant $A(\omega)= A_0$,   $\lambda (A) =  \lim_{\gamma} \lambda_{\gamma}(A)=\Tr[A_0]$. Thus, the limit functional $\lambda$ is still in the convex set $\mathscr C$, i.e. $\mathscr C$ is weak-* closed.  \qed  

A straightforward consequence of the previous two lemmas is the
following
\begin{Theo}[Compactness of $\mathscr C$]\label{Theo:Compact}
The embedding set $\mathscr C$  is compact in the weak-* topology.
\end{Theo}
\Proof  According to the celebrated Banach-Alaoglu Theorem\cite{Banach-Alaoglu}, the unit ball in a dual Banach space is compact in the weak-* topology. Since $\mathscr C$ is a closed subset of a compact set, it is automatically compact. \qed

\section{Realization of measurements with compact outcome space}

For compact $\Omega$, Lemma \ref{Lemma:IdentificationCompact}
identifies the set $\mathscr P$ of all POVMs with the compact subset
$\mathscr C \subseteq \mathcal V^*$. The fact that the
set of POVMs is compact allows us to get a decomposition of any given
POVM in terms of extremal POVMs. Indeed, this can be done be
exploiting the Krein-Milman Thoerem of convex analysis, which states
that any compact convex set coincides with the closure of the convex
hull of its extremal points, or, in an alternative formulation, that
any compact convex set coincides with the convex hull of the closure
of its extremal points (in this case, the closure and the convex hull
can be permuted \cite{KreinMilman}).

For the convex set $\mathscr P$, the first formulation of Krein-Milman Theorem implies that   any POVM $P \in \mathscr P$ can be arbitrarily well approximated by a POVM $\widetilde P$ which is a randomization of extremals, i.e.
\begin{equation}
\widetilde P = \int_{ Ex(\mathscr P)}  p(\d x) ~ E^{(x)}~,
\end{equation}  
where $x$ is the random
variable parametrizing the extremal POVMs $E^{(x)} \in  Ex(\mathscr P)$. Notice that this
approximation is in the sense of the weak-* topology, for any
given function $f(\omega)$ and for any given quantum state $\rho$ there exists a POVM $\widetilde P$ of the above form that approximates arbitrarily well the expectation value produced by $P$. 

The second formulation of Krein-Milman Thoerem leads instead to the
exact decomposition
\begin{equation}
P = \int_{X} p(\d x)~ E^{(x)}~,  
\end{equation}   
where now the integral runs over the set $X$ defined as the closure of the set of extremal POVMs, namely $X := \overline{Ex (\mathscr P)}$, and, accordingly, the POVM $E^{(x)}$ is either extremal or a limit of extremals. This is the formulation of the Theorem used in Ref. \cite{PRL}.  

In the next Subsection we will use the Krein-Milman Theorem to show
that any POVM $P
\in \mathscr P$ is a continuous randomization of POVMs with finite
support, namely POVMs that can be obtained as the post-processing of a
finite quantum measurement. 

We recall the definition of support   of a measure $\mu(\d \omega)$ as the set of all points $\omega\in\Omega$ such that $\mu (B)
>0$ for any open set $B$ containing $\omega$. The following Lemma, proved in  Ref. \cite{PRL}, characterizes the support of extremal POVMs. 
\begin{Lemma}[Support of extremal POVMs]\label{Lemma:SuppExt}
  Let $P\in \mathscr P$ be a POVM and $\mu(\d\omega)$ the measure defined by
  $\mu (B) = \Tr[P(B)] \quad \forall B \in \sigma (\Omega)$.  If $P$ is extremal, then the support of
  $\mu(\d\omega)$ is finite and contains no more than $d^2$ points.
\end{Lemma}
\Proof See Ref. \cite{PRL}.


Since extremal POVMs do have finite support (their support contains
less than $d^2$ points), it is enough now to prove that any weak-*
limit of extremal POVMs has a finite support.
\begin{Lemma}[Limits of extremal POVMs with compact outcome space]
Let $ P \in \mathscr P$ be a weak-* limit of extremal POVMs, namely $P =\lim_{\gamma} E_{\gamma}$ for some net $\{ E_{\gamma} \in Ex (\mathscr P)~|~ \gamma \in \Gamma\}$ of extremal POVMs. Then, $P$ has a finite support, containing no more than $d^2$ points.
\end{Lemma}
\Proof Due to Lemma \ref{Lemma:SuppExt},  for any $\gamma \in \Gamma$  the extremal POVM $E_{\gamma}$ can be written as 
\begin{equation}
E_{\gamma}(B) = \sum_{i=1}^{d^2} \chi_B \left (\omega^{(\gamma)}_i\right)~ P_i^{(\gamma)}~,
\end{equation}
where $\chi_B (\omega)$ is the characteristic function of the set
$B\subseteq \Omega$, the finite set $\{\omega_i \in \Omega~|~ i =1,
\dots , d^2\}$ is the support of the measure $\mu_{\gamma}(\d \omega)$
defined by $\mu_{\gamma}(B) = \Tr[E_{\gamma}(B)]$, and
$\{P_i^{\gamma}~|~ i= 1, \dots, d^2\}$ is a finite POVM.  Consider the
net of finite POVMs given by $\{ P^{(\gamma)}= \{P_i^{(\gamma)}\}~|~
\gamma \in \Gamma\}$. Since the set of POVMs is compact, one can find
a subnet $\{P^{(\gamma_0)}~|~ \gamma_0 \in \Gamma_0 \subseteq
\Gamma\}$ such that $\lim_{\gamma_0} P_i^{\gamma_0} = P_i \quad
\forall i$ for some finite POVM $P=\{P_i\}$. Morever, consider the net
of points $\{\omega_1^{(\gamma_0)}~|~ \gamma_0 \in \Gamma_0\}$. Since
$\Omega$ is compact, one can find a subnet $\{
\omega_1^{(\gamma_1)}~|~ \gamma_1 \in \Gamma_1 \subseteq \Gamma_0 \}$
such that $\lim_{\gamma_1} \omega^{(\gamma_1)}_1 = \omega_1$ for some $\omega_1 \in \Omega$. Iterating the procedure $N= d^2$ of times, one obtains a directed set $\Gamma_{N}$ along which the net of POVMs $\{P^{(\gamma_N)}\}$ and all the nets of points $\{\omega_i^{(\gamma_N)}\}$ converge. Taking the limit along the directed set $\Gamma_{N}$ we thus obtain 
\begin{eqnarray}
P(B) &=&  \lim_{\gamma_N}~ \left[ \sum_{i =1}^{d^2} ~ \chi_B \left (\omega^{(\gamma_{N})}_i\right)~ P_i^{(\gamma_{N})} \right]\\
&=&
\sum_{i =1}^{d^2} ~ \chi_B \left (\omega_i\right)~ P_i~.
\end{eqnarray}  
This proves that for compact outcome space $\Omega$ any  limit of extremal POVMs has finite support. The support is indeed contained in the set of points $\{\omega_i ~|~ i=1, \dots, d^2\}$. \qed

The Krein-Milman Theorem then provides the realization Theorem for POVMs with compact outcome space:

\begin{Theo}[Realization of POVMs with compact outcome space \cite{PRL}]\label{Theo:RealizationCompact}
Any POVM $P$ with compact outcome space $\Omega$ can be decomposed as
\begin{equation}\label{decompos}
P(B)=\int_{\mathcal{X}} p(\d x)~ E^{(x)}(B),
\end{equation}
where $x\in{\mathcal X}$ is a suitable random variable, $p(x)$ a probability density, and, for every value of
$x$, $E^{(x)}$ denotes a POVM with finite support, i.e. of the form 
\begin{equation}
\label{decomposE}
E(B)=\sum_{i=1}^{d^2}\chi_B(\omega_i) P_i
\end{equation}
$\{\omega_i \in \Omega\}$ being a set of points, and $\{P_i\}$ being a
finite POVM with at most $d^2$ outcomes.
\end{Theo}

\section{Realization of measurements with noncompact outcome space}
\subsection{Realization Theorem for arbitrary POVMs}

For noncompact $\Omega$, the problem is that the correspondence
between the set of POVMs $\mathscr P$ and the compact set $\mathscr C
\subseteq \mathcal V^*$ is not surjective, as explained in the Remark
after Lemma \ref{Lemma:IdentificationCompact}. Hence, the set
$\mathscr P$ cannot be identified with $\mathscr C$.  Nevertheless,
since $\mathscr C$ is compact, one can still use the Krein-Milman
Theorem to get a decomposition of any functional $\lambda \in \mathscr
C$ in terms of extremal functionals. This automatically provides a
decomposition for POVMs, since they form a proper subset of $\mathscr
C$. However, such a decomposition might contain extremal functionals
that do not correspond to POVMs.  In the following we will show that, luckily, this is not the case, namely the decomposition of any POVM $P \in
\mathscr P$ only contains extremal POVMs and POVMs that are limits of
extremals. In other words, the extremal functionals in $\mathscr C$
that are not in correspondence with POVMs do not show up in the
decomposition of any POVM.

In order to tackle the problem of noncompact outcome space $\Omega$, we will reduce it to a problem with compact outcome space, using the following
\begin{Lemma}[Compactification of the outcome space] \label{Lemma:Compactification}
The Banach space $\mathcal V$, containing
all bounded and continuous functions from $\Omega$ to $\Bndd{\spc
H}$, is isomorphic to the Banach space of continuous functions from
$\overline{\Omega}$ to $\Bndd{\spc H}$ where $\overline \Omega \supset \Omega$ is a suitable compact Hausdorff space. 
\end{Lemma}

\Proof The Lemma is a direct consequence of the well-known Gelfand isomorphism\cite{Gelfand}, by which the $C^*$-algebra of continuous bounded functions $\mathcal C_b (\Omega)$ is isomorphic to a $C^*$-algebra $\mathcal C (\overline \Omega)$ of continuous functions on a compact Hausdorff space $\overline \Omega \supset \Omega$.  Indeed,  the Banach space $\mathcal V$ is by definition the tensor product $\mathcal C_b (\Omega) \otimes \Bndd {\spc H}$. Then, the isomorphism $\mathcal C_b(\Omega) \cong \mathcal C(\overline \Omega)$ implies $\mathcal V \cong \mathcal C(\overline \Omega ) \otimes \Bndd{\spc H}$, i.e. the Banach space $\mathcal V$ is isomorphic to the Banach space of continuous functions from $\overline \Omega$ to $\Bndd{\spc H}$. \qed
The above Lemma, combined with Lemma \ref{Lemma:IdentificationCompact}, shows that the convex set $\mathscr C \subseteq \mathcal V^*$ can be identified with the auxiliary set of POVMs with compact outcome space $\overline \Omega$. In the following we will denote this auxiliary set as $\overline{\mathscr P}$ \cite{Note:PBar}. Clearly,  the set $\overline {\mathscr P}$ contains $\mathscr P$ as a proper subset.  

Since  $\overline{\Omega}$ is compact, we can use Theorem \ref{Theo:RealizationCompact} to decompose any POVM in $\overline {\mathscr P}$, thus getting the following 
\begin{Lemma}[Realization of POVMs in the auxiliary set $\overline {\mathscr P}$]\label{Lemma:RealizationAux}
Any POVM $P \in \overline {\mathscr P}$ can be decomposed
as
\begin{equation}\label{decomposAux}
P(B)=\int_{\mathcal{X}} p(\d x)~ E^{(x)}(B),
\end{equation}
where $x\in{\mathcal X}$ is a suitable random variable, $p(x)$ a probability density, and, for every value of
$x$, $E^{(x)} \in \overline {\mathscr P}$ denotes a POVM with finite support, i.e. of the form 
\begin{equation}
\label{decomposEAux}
E(B)=\sum_{i=1}^{d^2}\chi_B(\bar \omega_i) P_i
\end{equation}
$\{\bar \omega_i \in \overline \Omega\}$ being a set of points, and $\{P_i\}$ being a
finite POVM with at most $d^2$ outcomes.
\end{Lemma}

Now we want to use the above Lemma to get a decomposition of the POVMs
with noncompact outcome space $\Omega$, exploiting the fact the
$\mathscr P$ is a proper subset of $\overline {\mathscr P}$. To this purpose,
we  prove that for a POVM $P \in \mathscr P$ the decomposition
of Eq. \eqref{decomposAux} only contains POVMs $E^{(x)}$ with support
in $\Omega$. In other words, we  show that the POVMs $E^{(x)}$
with support containing points of $\overline \Omega- \Omega$ have zero
measure in the decomposition of any $P \in \mathscr P$. 

In the following we will call \emph{regular} any POVM in $\overline {\mathscr P}$ such that $P(\Omega) = I$, and \emph{singular} a POVM such that $P(\Omega) < I$ (since we are in the auxiliary set $\overline {\mathscr P}$, the POVM $P$ has only to be normalized on $\overline \Omega$, i.e.  $P (\overline \Omega) = I$). Of course, all POVMs in $\mathscr P$ are regular, due to their normalization. On the other hand, a finitely-supported POVM $E$, of the form $E(B) = \sum_{i=1}^{d^2} \chi_B (\bar \omega_i)$, is regular if all  points $\{\bar \omega_i~| i = 1,\dots, d^2\}$ are in $\Omega$, while it is singular if some point $\omega_i$ is in $\overline \Omega -\Omega$.  


Using these observations, we finally obtain the desired decomposition Theorem:
\begin{Theo}[Realization of POVMs with noncompact outcome space \cite{PRL}]\label{Theo:RealizationNoncompact}
Any POVM $P$ with noncompact outcome space $\Omega$ can be decomposed as
\begin{equation}
P(B)=\int_{\mathcal{X}} p(\d x)~ E^{(x)}(B),
\end{equation}
where $x\in{\mathcal X}$ is a suitable random variable, $p(x)$ a probability density, and, for every value of
$x$, $E^{(x)}$ denotes a POVM with finite support, i.e. of the form 
\begin{equation}
E(B)=\sum_{i=1}^{d^2}\chi_B(\omega_i) P_i
\end{equation}
$\{ \omega_i \in  \Omega\}$ being a set of points, and $\{P_i\}$ being a
finite POVM with at most $d^2$ outcomes.
\end{Theo}
\Proof 

Consider $P$ as an element of $\overline {\mathscr P}$, take the its
decomposition as in Lemma \ref{Lemma:RealizationAux}, and
separate in the integral the two contributions of regular POVMs and
singular ones, by splitting the set $\mathcal X$ into the two disjoint
regions $\mathcal X_{reg}$ and $\mathcal X_{sing}$. In this way, one can write
$P = \alpha P_{reg} + (1-\alpha) P_{sing}$, where $\alpha$ is given by
\begin{equation}
\alpha = \int_{\mathcal X_{reg}} p(\d x)~, 
\end{equation} 
the regular part $P_{reg}$ of the POVM $P$ is given by
\begin{equation}
P_{reg} (B) = \left\{ 
\begin{array}{lr}
\frac 1 {\alpha}~ \int_{\mathcal{X}_{reg}} p(\d x)~ E^{(x)}(B)~,\qquad & \alpha \not =0\phantom{~,}\\
0~, & \alpha =0~,
\end{array}
\right.
\end{equation}
and  the singular part is given by
\begin{equation}
P_{sing} (B) =
\left\{ 
\begin{array}{lr}
\frac 1 {(1-\alpha)}~ \int_{\mathcal{X}_{sing}} p(\d x)~ E^{(x)}(B)~,\qquad & \alpha \not =1\phantom{~.}\\
0~, & \alpha =1~.
\end{array}
\right.
\end{equation}
Now, since the POVM $P$ is a POVM on $\Omega$ we have 
\begin{eqnarray} 
\openone &=& P(\Omega)\\
&=&  \alpha    P_{reg}(\Omega) + (1-\alpha) P_{sing}(\Omega)\\ 
&\le& \openone~,
\end{eqnarray} 
where we used that $P_{reg} (\Omega) = \openone$, while $P_{sing} (\Omega) < \openone$.  In order to satisfy the equality, one must necessarily have $\alpha =1$, namely, the singular POVMs  have zero measure in the decomposition of $P$. \qed


\begin{thebibliography}{50}
\bibitem{Note:PBar} The notation $\overline P$ is consistent with the fact that the set $\overline {\mathscr P}$ is the weak-* closure of the set of POVMs $\mathscr P$.
\end{thebibliography}
 \end{document}